\documentclass[12pt,preprint]{aastex}

\usepackage{lscape}
\usepackage{rotating}
\usepackage{rotfloat}
\usepackage{epsfig}
\usepackage{natbib}

\newcommand{\sax}{{\it BeppoSAX} }

\newcommand{\integral}{{\it INTEGRAL}}
\newcommand{\swift}{{\it SWIFT} }

 \def\source{IGR~J16328-4726}
\def\be{\begin{equation}}
\def\ee{\end{equation}}

\shorttitle{\source\/}
\shortauthors{Fiocchi et al.}

\begin{document}


\title{The INTEGRAL source \source\/: a High Mass X-ray Binary from the \sax\/ era}

\author{M. Fiocchi\altaffilmark{1}, A. Bazzano\altaffilmark{1}, A. J. Bird\altaffilmark{3}, S. P. Drave\altaffilmark{3},  L. Natalucci\altaffilmark{1}, P. Persi \altaffilmark{1}, L. Piro\altaffilmark{1, 2}, P. Ubertini \altaffilmark{1} }

\altaffiltext{1}{Istituto di Astrofisica e Planetologia Spaziali (INAF). Via Fosso del Cavaliere 100, Roma, I-00133, Italy}
\altaffiltext{2}{Astronomy Department, Faculty of Science, King Abdulaziz University, P.O. Box 80203, Jeddah 21589, Saudi Arabia}
\altaffiltext{3}{School of Physics and Astronomy, University of Southampton, University Road, Southampton SO17 1BJ, UK }

\begin{abstract}
We report on  temporal and spectral analysis of the INTEGRAL fast transient candidate \source\/ observed with \sax\/ in 1998 and more recently with INTEGRAL.
The MECS X--ray data show a frequent micro activity typical of the intermediate state of Supergiant Fast X-ray Transients and a weak flare with duration of $\sim$4.6 ks.
The X--ray emission in the 1.5--10 keV energy range is well described through the different time intervals 
by an absorbed power law model. Comparing spectra from the lower emission level up to the peak of the flare, we note that  
while the power-law photon index was constant ($\sim$ 2), the absorption column density varied by a factor of up to $\sim$6-7, reaching the value of  
$\sim$2$\times$10$^{23}$~cm$^{-2}$  at the peak of the flare.\\
Analysis of the long-term INTEGRAL/IBIS light curve confirms and refines the proposed $\sim$10.07 day period, and the derived ephemeris places the \sax\/ observations away from periastron.\\
Using the near and the mid-IR available observations, we constructed a spectral infrared distribution for the counterpart of  \source\/, allowing us to identify its counterpart as a High Mass OB type star, and to classify this source as a firm HMXB.\\
Following the standard clumpy wind theory, we estimated the mass and the radius of the clump responsible of the flare. The obtained values of M$_{cl}\simeq4\times10^{22}g$ and R$_{cl}\simeq4.4\times10^6$~km are in agreement with expected values from theoretical predictions.

\end{abstract}


\section{Introduction}
The X-ray fast transient source IGR J16328-4726 was discovered by Bodaghee et al. (2007).
Fiocchi et al. 2010, making use of the IBIS/INTEGRAL data, tentatively classified it as a  Supergiant Fast X--ray Transient (SFXT) candidate on the basis of
the transient and recurrent nature, the spectral properties observed
during the outbursts that lasted only a few tens of minutes/hours, the location in the Galactic Plane, the faint quiescent emission ($<2.5\times10^{-12}~erg~s^{-1}~cm^{-2}$) and a flux dynamic range $>$170 in the 0.3-10 keV (XRT/Swift) and $>$165 in the 20-50 keV (IBIS/INTEGRAL) bands.
The source characteristics and behavior  are reported in Fiocchi et al. (2010) and references therein. An additional detection was reported in the Swift BAT 54 month all-sky survey (Cusumano et al. 2010) with an averaged flux of $(3.2\pm0.8)\times~10^{-11}~erg~cm^{-2}~s^{-1}$ in the 15-70 keV energy band.
\begin{figure*}[h]
\begin{center}
\includegraphics[width=0.43\textwidth]{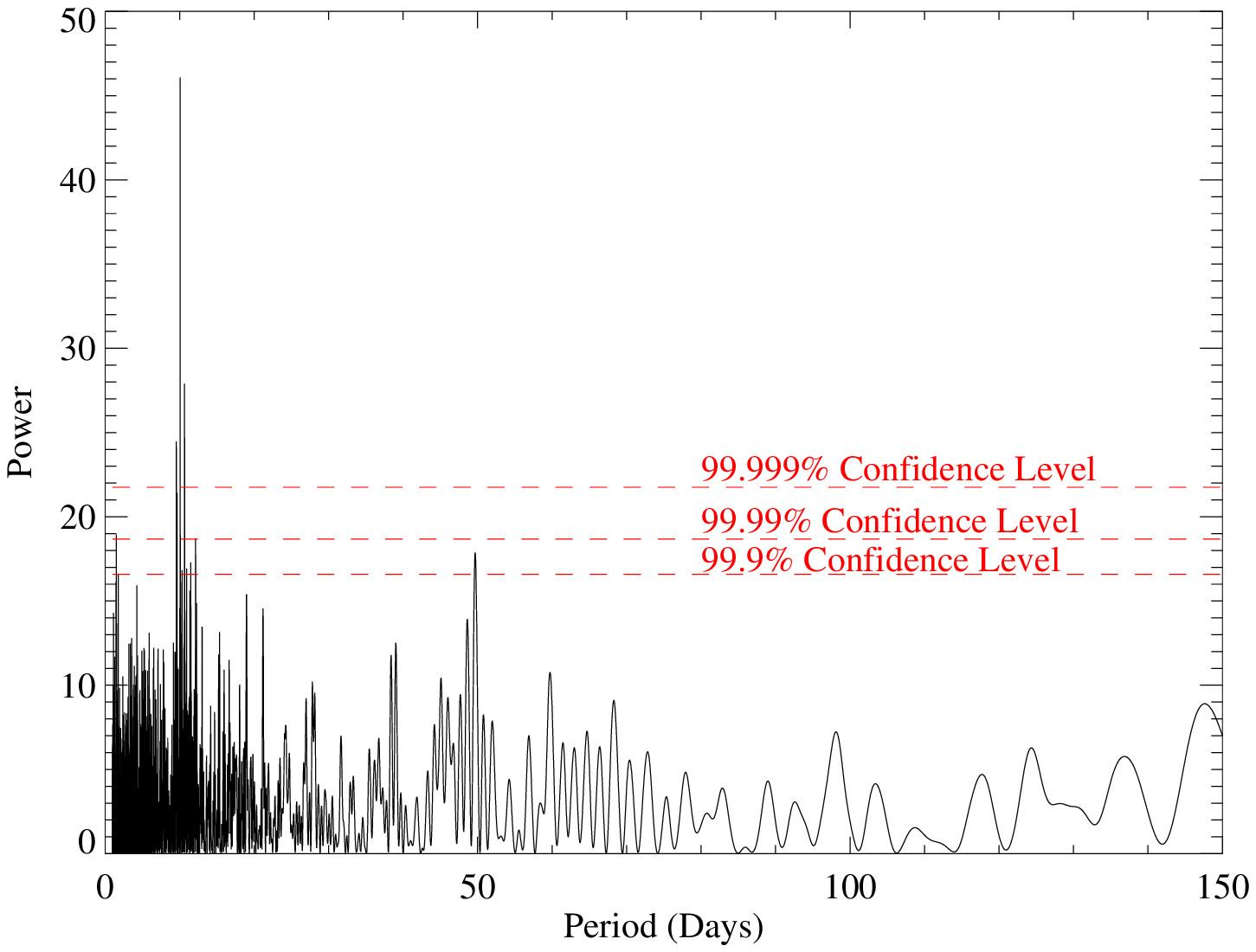}
\hspace{1cm}
\includegraphics[width=0.43\textwidth]{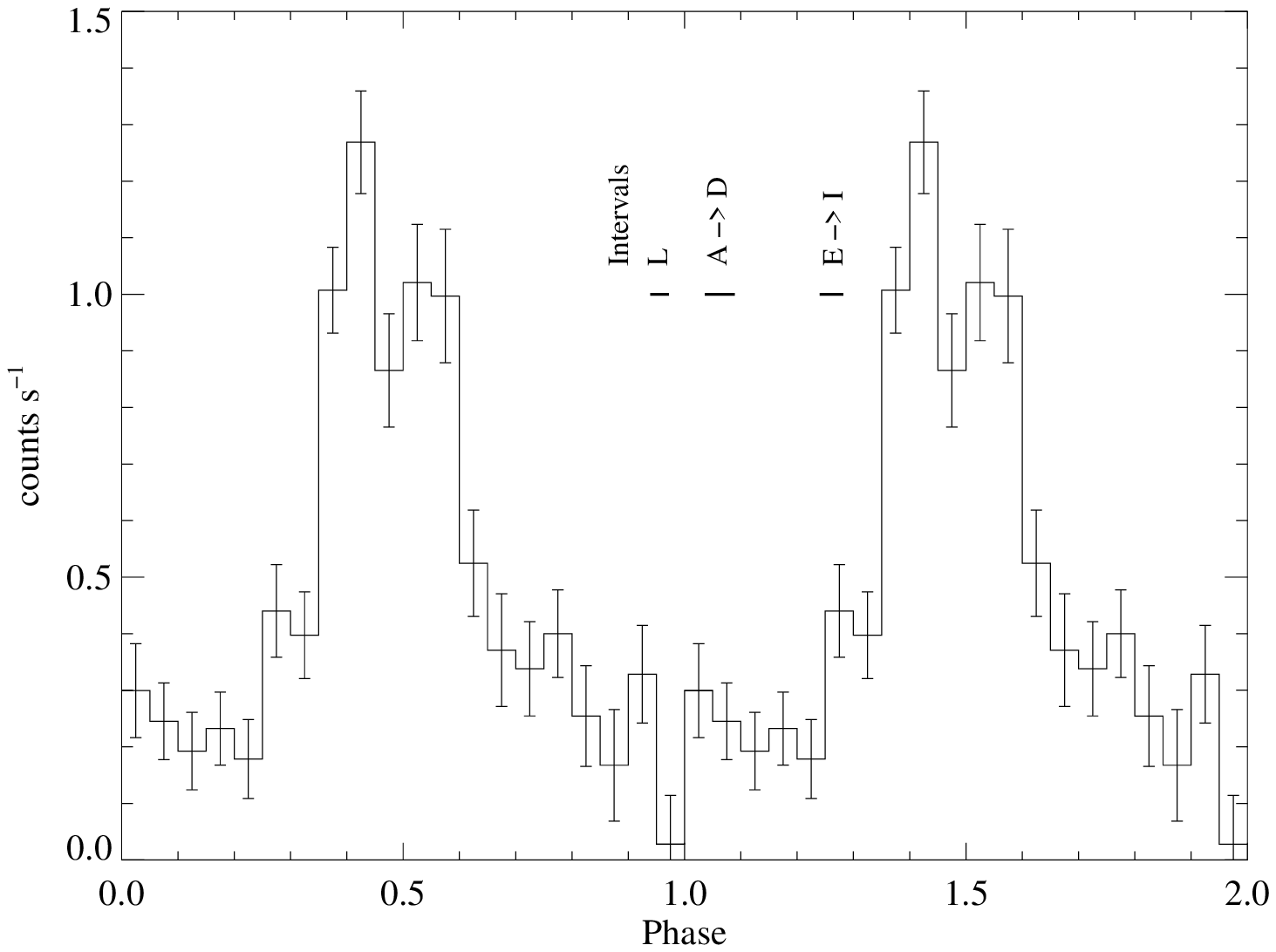}
\caption{Left: Lomb-Scargle periodogram of the 18$-$60\,keV IBIS/ISGRI light curve of \source\/ showing the detection of the orbital period at 10.068$\pm$0.002\,days. Right: IBIS/ISGRI 18$-$60\,keV phase folded light curve using the best orbital period determination of 10.068\,days and a zero phase ephemeris of MJD 52651.164. The locations of the \sax\/ observations within orbital phase using the stated ephemeris are also indicated.}
\label{fig:IBISdata}
\end{center}
\end{figure*}

The physical mechanism producing short X-ray flares of the SFXTs is still under discussion, and several scenarios have been proposed. The most frequently discussed scenario is the clumpy stellar wind model (in't Zand 2005; Walter
\& Zurita Heras 2007) where dense clumps in an isotropic stellar wind cause short X-ray flares
when they accrete onto a neutron star. A few SFXT systems can be explained by an eccentric NS orbit around a massive star
with a disk-like anisotropic stellar wind (Sidoli et al. 2007).
The SFXT emission could be explained also by a gating effect caused by the extremely strong NS magnetic field ($\sim10^{14}$ G) and rotational properties (Bozzo et al. 2008a). 

Proximity to the Black Hole Candidate 4U~1630-47, has previously made any periodicity study based on long-term hard X-ray light curves difficult, but 
Corbet et al. (2010) reported a significant period at $\sim 10$ days from \source. 

Here we report on analysis of \sax observations performed in 1998, a re-analysis of the INTEGRAL/IBIS light curve, and an archival search for other observations that might allow us to confirm or discard the SFXT nature of \source\/.

\begin{figure}
\begin{center}
\label{fig:lcim}
\includegraphics[totalheight=195pt,width=420pt]{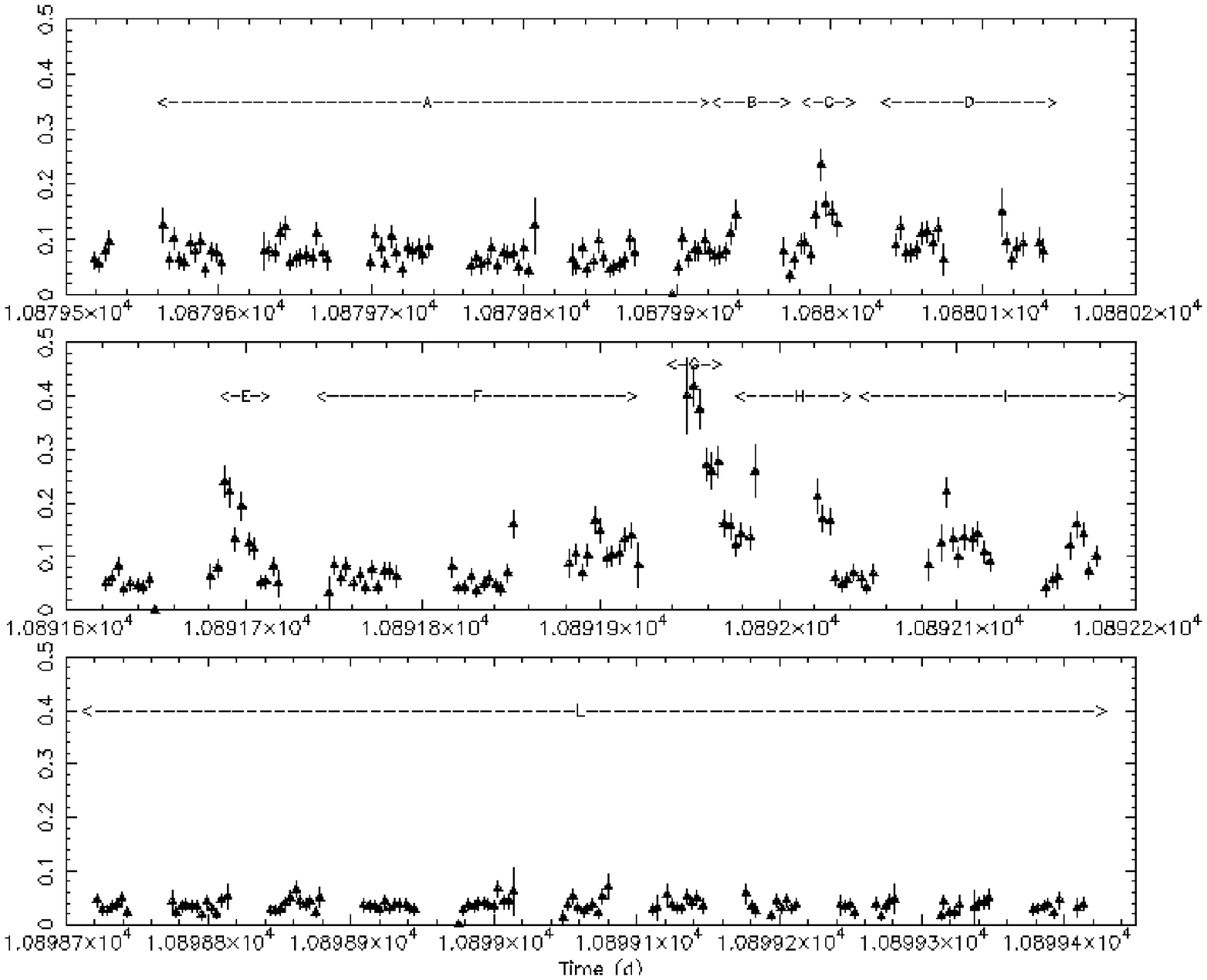}
\vspace{1cm}
\includegraphics[angle=-90,totalheight=200pt,width=200pt]{fig1a.eps}
\caption{Top: the MECS light curve of \source\/ in the 1.5--10 keV energy band with bin size 300 s, using \sax\/ observations performed in 1998. Bottom: the MECS mosaic image in the 1.5--10 keV energy band during Phase G reported in the top panel. The white circle shows \source\/ position and the 4' radius circle is the region from which we have extracted the spectra. The bright source is the nearby BH candidate 4U1630-47. }
\end{center}
\end{figure}

\section{INTEGRAL observations and data analysis}

The INTEGRAL dataset used included all public data available for the period up to 30$^{th}$ September 2010 (MJD~52650 through MJD~55469). The total on-source exposure on \source\/ during this period was $\sim$9.2\,Ms. Light curves for both \source\/ and  4U~1630-47 were extracted from science window (pointing) images constructed using the standard OSA9 analysis software, in the 18-60 keV energy band. The light curves were then filtered to remove science windows (ScW) with an exposure of less than 200\,s and/or in which the sources were at an off-axis angle of greater than 12\,degrees as such ScWs have large intrinsic uncertainties in flux measurement. The final, filtered \source\/ light curve has an on source exposure of $\sim$7.1\,Ms. We performed a timing analysis on the derived \source\/ light curve using standard Lomb-Scargle techniques (Lomb 1986, Scargle 1982) to search for periodic signals in the data, and monte-carlo randomisation to determine both the significance of, and uncertainty in, the derived periodicities (for further information on the randomisation techniques applied see Drave et al. 2010 and references thererin).
\begin{figure}[h]
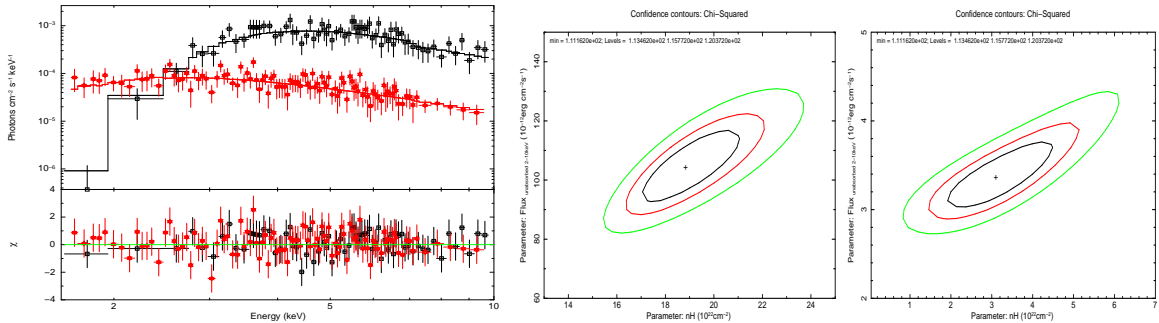

\begin{center}
\includegraphics[angle=-90,totalheight=120pt,width=190pt]{fig2a.eps}
\includegraphics[angle=-90,totalheight=120pt,width=120pt]{fig2b.eps}
\includegraphics[angle=-90,totalheight=120pt,width=120pt]{fig2c.eps}
\caption{Left: the MECS spectra shown together with the power law models at the peak emission (Phase G) and at the low emission level (Phase L) are indicated with squares and stars, respectively. Middle: Contour level of the column density versus the unabsorbed flux in the 2-10 keV measured from the spectrum during Phase G. The contours correspond to the 68\%, 90\% and 99\% confidence level. Right: Contour level of the column density versus the unabsorbed flux in the 2-10 keV measured from the spectrum during Phase L. The contours correspond to the 68\%, 90\% and 99\% confidence level. 
}
\label{fig:contours}
\end{center}
\end{figure}

\section{INTEGRAL timing analysis}

The Lomb-Scargle periodogram of the 18$-$60\,keV IBIS/ISGRI light curve of \source\/ (Figure~\ref{fig:IBISdata}, left panel) shows the detection of a period at 10.068 days, which we interpret as the likely orbital period. A randomisation monte-carlo determined the 1-sigma uncertainty on that period to be 10.068$\pm$0.002\,days. This period is consistent, within errors, with that reported in Corbet et al. (2010) and provides an independent confirmation of the orbital period of the binary system. The detection in that work used a \swift\//BAT data set spanning MJD~53359 through MJD~55104. This data set began after the last period of high X-ray activity in 4U~1630-47 ceased at $\sim$MJD~53300. The \integral\//IBIS data set used in this work does span the earlier region of high activity of 4U~1630-47. However there is only a small amount of contamination from this bright source in the \source\/ light curve and a comparative analysis performed using only data from the same MJD range as Corbet et al. (2010) provided a consistent detection of the period. The higher exposure in the full data set allowed a more accurate determination of the period and its uncertainty however and thererfore it is this data set that is used in the remainder of this work.  

Folding the light curve using the best orbital period determination of 10.068\,days and a zero phase ephemeris, taken as the first \integral\/ observation in the light curve, of MJD 52651.164 yields the folded light curve shown in Figure~\ref{fig:IBISdata} (right panel). This shows a single broad emission maximum above a quiescent level, but the statistical quality makes it difficult to derive further information. The shape of the profile is similar to those observed from other SFXTs such as IGR~J17544-2619 (Clark et al. 2009) and IGR~J16465-4507 (Clark et al. 2010). The times of the \sax\/ observations within this orbital phase using the stated ephemeris are also indicated.
\begin{figure*}[h]
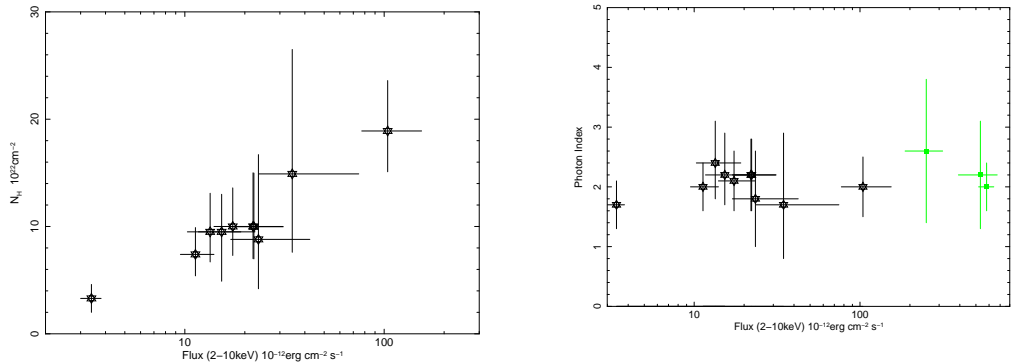

\begin{center}
\includegraphics[angle=-90, width=0.38\linewidth]{fig3a.eps}
\hspace{1cm}
\includegraphics[angle=-90, width=0.35\linewidth]{fig3b.eps}
\caption{Left: The column density versus the unabsorbed flux in the 2--10 keV energy range, for each studied period reported in figure 1.
Right: The photon index versus the unabsorbed flux in the 2--10 keV energy range, for each studied period reported in figure 1. The green points are the photon index versus the unabsorbed flux in the 20--50 keV energy range
using the IBIS/INTEGRAL data (from Fiocchi et al. 2010).
}
\label{fig:parameters}
\end{center}
\end{figure*}

\section{\sax\/ observations and data analysis}

Results presented here are from the Low-Energy Concentrator Spectrometer (LECS; Parmar et al. 1997) and the Medium-Energy Concentrator Spectrometer (MECS; Boella et al. 1997) on-board BeppoSAX.
The High Pressure Gas Scintillation
Proportional Counter (HPGSPC; Manzo et al. 1997) and the Phoswich Detection System (PDS; Frontera et al. 1997) instruments were not considered as their field of view included also emission from the
nearby Black Hole Candidate 4U1630-47. \\ 
The LECS and MECS event files and spectra were generated with the Supervised Standard Science Analysis (Fiore, Guainazzi \& Grandi 1999).  
The MECS spectra were accumulated in circular regions of $4\arcmin$ radius and publicly available response matrices were used. The off axis corrections have been taken into account using  the apposite response matrices.\\
The Point Spread Function (PSF) of the MECS instruments has been modeled as the sum of 2 components: a Gaussian describing mainly the detector PSF and a King Profile describing mainly the optics PSF (Boella et al. 1997). The final result is a fully analytic differential PSF with a width of 2.5' at 6.4 keV, 80\% of encircled energy radius. Since the distance between the Black Hole 4U 1630-47 and IGR J16328-4726 is $\sim 14^{\prime}$, the Black Hole contribution within the spectral extraction radius of the IGR J16328-4726 is widely below 1\% (being this the systematic error added to the data following the indications given in Fiore, Guainazzi \& Grandi (1999) for the standard spectral analysis) .\\
In the archival MECS images (1.5--10 keV) \source\/ was detected above 5 sigma during three different observations performed on 1998-03-07 12:23:21.0 UTC, 1998-03-19 14:52:11.0 UTC and 1998-03-26 17:16:48.5 UTC. 
The LECS image analysis (0.5--3.5 keV) showed that \source\/ was not detected above 5 sigma during these three observations. 
Log of these observations are shown in Table 1, reporting start times, exposure times and the \source\/ count rate for each observations.

\section{The \sax\/MECS results}

Figure 2 (upper panel) shows the MECS light curve of \source\/ with a bin size of 300 s.
As previously observed by IBIS on board of INTEGRAL satellite (Fiocchi et al. 2010), the source shows
highly recurrent micro-activity with durations from tens of minutes to several hours.

To search for spectral changes, we accumulated X-ray spectra during several time intervals, as indicated on the light curve. The X-ray spectra are all well fitted by an absorbed power-law in the 1.5--10 keV energy range. Details of the time intervals, and results of this analysis for each selected period are reported in Table 2.
\begin{figure*}[h]
\begin{center}
\fbox{\includegraphics[width=0.45\linewidth]{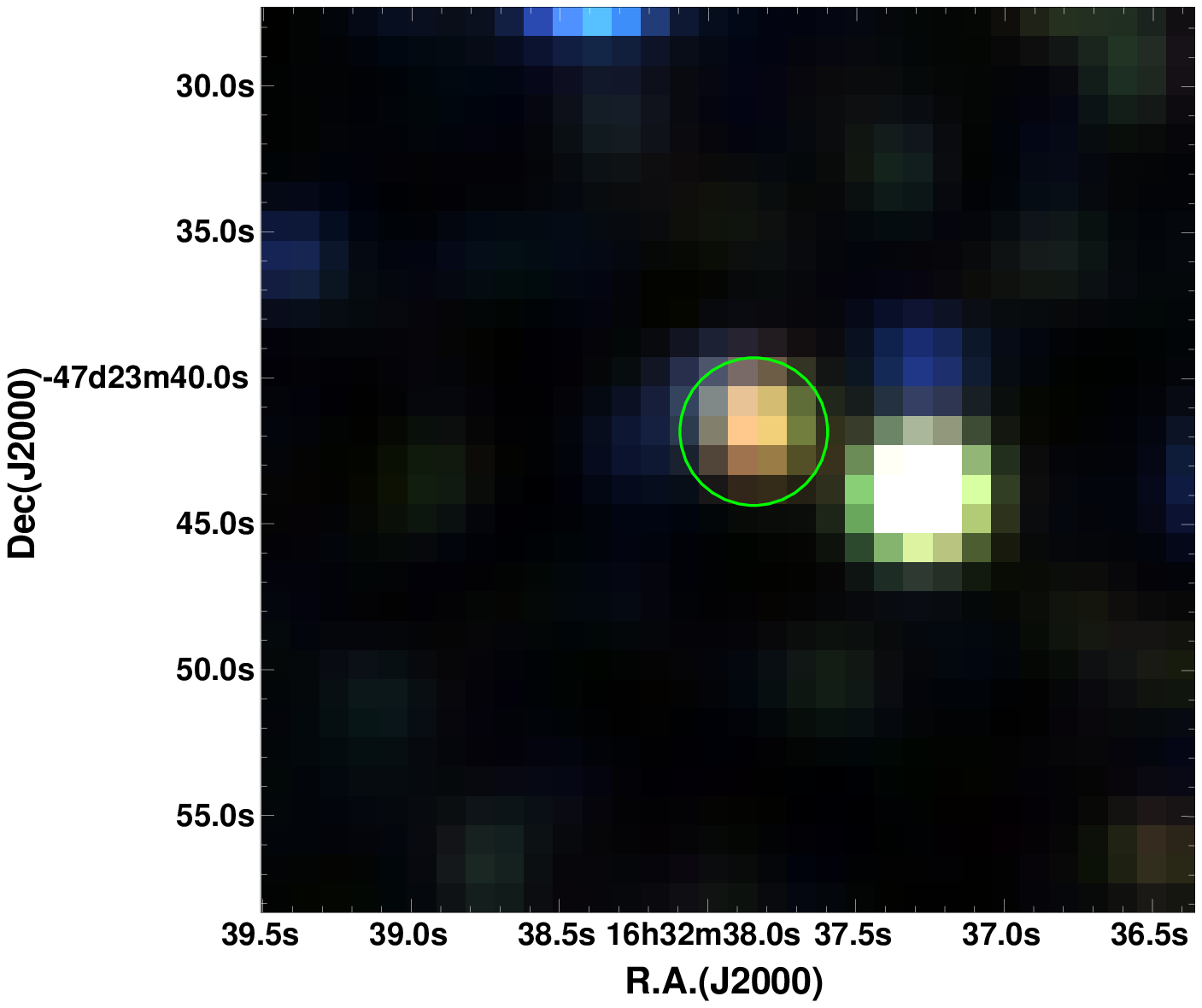}}
\hspace{1cm}
\fbox{\includegraphics[width=0.35\linewidth]{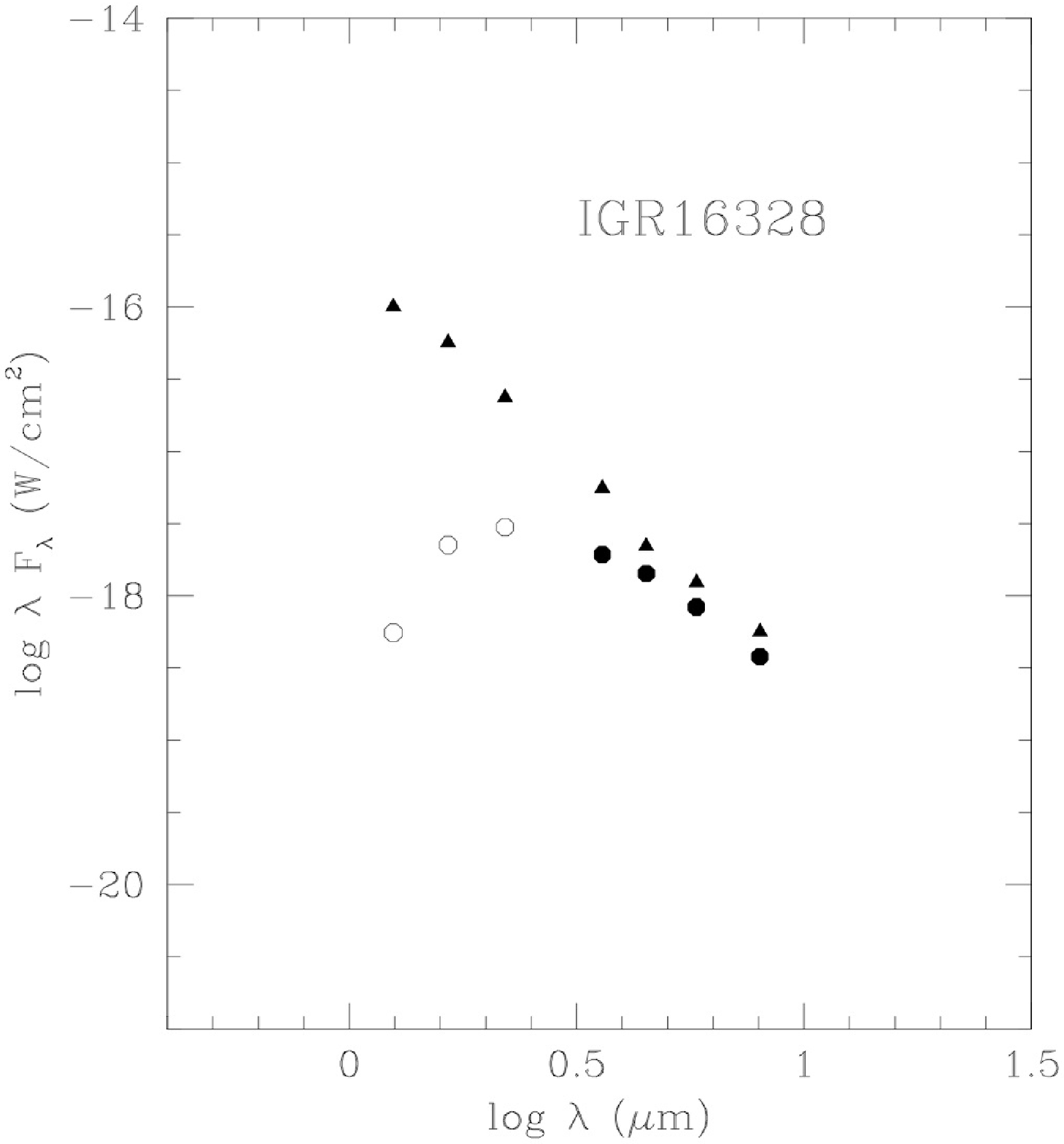}}
\caption{Left panel: Color-coded 2MASS image centered on \source\/ made  from J(blue), H(green) and K(red) individual images. The circle is the XRT uncertainty at 99\% confidence level from Grupe et al. 2009.  Right panel: the infrared energy distribution of the \source\/ counterpart.
}
\label{figIR}
\end{center}
\end{figure*}

Figure ~\ref{fig:contours} (left panel) shows the MECS spectra and residuals with respect to the absorbed power law models at the peak of emission (phase G) and at the lowest emission level (phase L), indicated with squares and stars, respectively. 
A plot of the evolution of the spectral parameters of the source is provided in Fig.~\ref{fig:parameters}.  
The power-law photon index remained fairly constant (see Fig.~\ref{fig:parameters}, right panel) including both the MECS and IBIS data from Fiocchi et al. 2010,
whereas the absorption column density underwent changes (Fig.~\ref{fig:parameters}, left panel). 
In particular, we measured a large increase (factor of $\sim$6-7) in the absorption column density 
from the lower emission level (phase L) up to the peak of the flare (period G) when the column density reached a value of  
$\sim$2$\times$10$^{23}$~cm$^{-2}$. 
To test the significance of the measured variation 
in the absorption column density, we report in Fig.~\ref{fig:contours} the $N_{\rm H}$-Flux 
parameters confidence contours for the relevant spectra in phases G and L (middle and right panel).   
The contours correspond to 68\%, 90\%, and 99\% confidence level. A very significant 
increase in the absorption column density with the unabsorbed fluxes is clearly detected. 
Although adding an iron line does not significantly improved the fit, we considered this component to estimate the iron line equivalent width upper limits.
With the iron line centroid fixed at 6.5 keV, we obtained equivalent width upper limits of 440 eV and 262 eV (3$\sigma$ confidence level) for the peak of the flare (period G) and the lower emission level (phase L), respectively. 
\begin{table*}
\begin{center}
\caption{BeppoSAX Observations}
\label{jou}
\scriptsize
\begin{tabular}{ccccc}
\hline\hline
& & & & \\
{\bf Start Time}&\multicolumn{2}{c}{\bf Exposure time}&\multicolumn{2}{c}{\bf Count s$^{-2}$}\\
& & & & \\
\hline
& & & & \\
&LECS&MECS&LECS&MECS\\
UTC&s&s&[0.5-3.5 keV]&[1.5-10 keV]\\
& & & & \\
\hline
& & & & \\
1998-03-07 12:23:21.0 &12290 &  29530  &   n.d.$^{a}$   &$7.3\pm0.2$   \\
1998-03-19 14:52:11.0 &9912   &  27786  &   n.d.$^{a}$   &$10.0\pm0.2$   \\
1998-03-26 17:16:48.5 &11694 &  31265  &   n.d $^{a}$   &$2.8\pm0.1$ \\
\hline
\hline
\end{tabular}\\
$^{a}$ n.d. means  that  the \source\/ was not detected above 5 $\sigma$. 
\end{center}
\end{table*}

\begin{table*}
\caption{Best-fit parameters of \source\/ during different time intervals 
of the \sax\/ observations (see Fig.~1)}
\tiny
\centering
\begin{tabular}{ccccccccccc}
\hline
\hline
\noalign{\smallskip} 
Interval & OBS. Tstart & Interval Tstart & Tstart Phase	& Exp.   & Rates 	& $N_{\rm H}$ 		& $\Gamma$       	&  $F_{\rm unabs}$                   	& $\chi^2$/d.o.f. \\
\noalign{\smallskip}
         & (s)    	&MJD& & (s)  &($10^{-2}$c/s) & (10$^{22}$~cm$^{-2}$) &                	& (10$^{-12}$erg~cm$^{-2}$~s$^{-1}$) 	&                           \\
\hline
\noalign{\smallskip} 
A 	& 1998-03-07  &  50879.563  &  0.036  &  1521.2 & 7.5$\pm$0.2 	& 7.7$\pm$1.5 		& 2.0$\pm$0.3    	&  12.7$^{+2.0}_{-1.5}$ 			& 80/105 			\\
	& 12:23:21.0  &             &    &    &             	&        		&                	&   									\\
\noalign{\smallskip}
B 	& 1998-03-07  &  50879.901 &  0.070  &  4713.5 & 7.1$\pm$0.4 	&  9.5$^{+3.6}_{-2.8}$ 	& 2.4$^{+0.7}_{-0.6}$ 	&  13.4$^{+5.7}_{-3.1}$ 		& 42/54 			\\
	& 12:23:21.0  &             &    &    &             	&        		&                	&   									\\
\noalign{\smallskip}
C 	& 1998-03-07  &  50879.988 &  0.079  &  1582.1 & 14.9$\pm$0.9 	& 14.9$^{+11.6}_{-7.3}$ & 1.7$^{+1.2}_{-0.9}$   &  34.6$^{+40}_{-11}$ 			& 29/36 			\\
	& 12:23:21.0  &            &    &    &             	&        		&                	&   									\\
\noalign{\smallskip}
D 	& 1998-03-07  &  50880.041 &  0.084  &  4308.9 & 8.2$\pm$0.4 	& 9.5$^{+3.5}_{-4.6}$ &2.2$^{+0.7}_{-0.5}$&  15.3$^{+7.2}_{-3.6}$ 				& 29/36 			\\
	& 12:23:21.0  &            &    &    &             	&        		&                	&   									\\
\noalign{\smallskip}
E 	& 1998-03-19  &  50891.681 &  0.240  &  3210.9 & 11.1$\pm$0.6	& 10$^{+5}_{-3}$ 	& 2.2$\pm$0.6    	&  22$^{+9}_{-5}$ 					& 31/53 			\\
	& 14:52:11.0  &            &    &    &             	&        		&                	&   									\\
\noalign{\smallskip}
F 	& 1998-03-19  &  50891.748 &  0.247  &  9577.5  & 7.0$\pm$0.3 	&  7.4$^{+2.5}_{-2.0}$ 	& 2.0$\pm$0.4 	 	&  11.3$^{+2.7}_{-1.8}$ 		& 69/93 			\\
	& 14:52:11.0  &            &    &    &             	&        		&                	&   									\\
\noalign{\smallskip}
G 	& 1998-03-19  &  50891.949 &  0.267  &  1614.9 & 30.1$\pm$1.4 	&  18.9$^{+4.7}_{-3.8}$ & 2.0$\pm$0.5    	&  104$^{+50}_{-27}$ 				& 42/67 			\\
	& 14:52:11.0  &            &    &    &             	&        		&                	&   										\\
\noalign{\smallskip}
H 	& 1998-03-19  &  50891.968 &  0.269  &  1569.6 & 14.0$\pm$0.8 	& 8.8$^{+7.9}_{-4.6}$	& 1.8$\pm$0.8    	&  23.4$^{+19.0}_{-6.45}$ 			& 33/33			 	\\
	& 14:52:11.0  &            &    &    &             	&        		&                	&   									\\
\noalign{\smallskip}
I 	& 1998-03-19  &  50892.029 &  0.274  &  6534.1  & 8.9$\pm$0.4 	& 10.0$^{+3.6}_{-2.7}$	& 2.1$\pm$0.5    	&  17.4$^{+5.8}_{-3.4}$ 			& 60/85			 	\\
	& 14:52:11.0  &            &    &    &             	&        		&                	&   									\\
\noalign{\smallskip}
L 	& 1998-03-26  &  50898.720 &  0.939  &  28460.6 & 2.8$\pm$0.1 	& 3.3$\pm$1.3 		& 1.7$\pm$0.3    	&  3.4$\pm0.4$ 				& 70/101 					\\
	& 17:16:48.5  &            &    &    &             	&        		&                	&   									\\
\noalign{\smallskip}
\hline
\hline
\noalign{\smallskip}
\end{tabular}\\
\scriptsize
The continuum spectral model is an absorbed power law. Here, $N_{\rm H}$ is the absorption column density, 
$\Gamma$ is the power-law photon index, $F_{\rm unabs}$ is the unabsorbed flux in the 1.5-10~keV band. 
In the last column we report the value of the $\chi^2$/d.o.f
for each spectrum. Exp. is the good time interval for the spectral extraction.\\
\normalsize
\label{tab:fit2}
\end{table*}

\section{Summary of the main outburst characteristics}
 The MECS light curve of \source\/ (Figure 2) shows
highly recurrent micro-activity with durations of tens of minutes and a strong outburst 
(its peak is reported as epoch G in Figure 2).
During the outburst detected with the MECS instrument, the source was not significantly detected ($>5\sigma$) in the corresponding LECS image, that had an exposure time of 775 s.  \\
The spectral analysis of the outburst peak  was reported in Table 2 (period G) for comparison with other activity period of the \source\/ and summarised in Table 3,
where we reported start time, duration and spectral parameters.  Using these values and assuming a distance of 3 kpc and 10 kpc, the X-ray luminosity (2--10 keV) spans from 10$^{35}$ erg s$^{-1}$ to 10$^{36}$ erg s$^{-1}$,  respectively.

To determine the best position, and as the source is highly absorbed, we extracted the MECS image in the 4--10 keV energy band during its outburst (G period reported in Figure~2). 
\source\/ was detected at $(51.5\pm3.0)\times10^{-2}$ count/s at RA(J2000)= 16h32m38.1s and Dec(J2000)=-47$^{\circ}$ 23' 56.8'' with a error box of 35 arcsec 
(Figure~2, lower panel).
Note that the circle reported in this figure is the radius for spectral extraction and not the position error box.

\section{Infrared identification}
The transient X-ray source \source\/ has been identified with the near-IR source 2MASS16323791-4723409 (Grupe et al. 2009), shown in Fig.~\ref{figIR} (left panel). The 2MASS catalogue reports only the J magnitude and gives an upper limit in H and K. From the images we have obtained the complete photometry of this source using the calibration reported in the headers. The measured magnitudes are  J=14.32$\pm$0.06, H=12.29$\pm$0.03 and K=11.19$\pm$0.02. The source  is reported in the DENIS survey as J1632379-472341 and has been detected also by the Spitzer/GLIMPSE survey. The magnitudes reported in the 4 IRAC bands obtained from the GLIMSPE catalogue (G336.7492+00.4223) are
M$_{3.6\mu m}$=10.21$\pm$0.05, M$_{4.5\mu m}$=9.81$\pm$0.06, M$_{5.8\mu m}$=9.63$\pm$0.05 and M$_{8.0\mu m}$=9.52$\pm$0.04.
From the observed near-IR colors of the source J-H=$2.03\pm0.06$ and H-K=$1.10\pm0.03$, we derive a visual extinction  of A$_V$=20. 
The source is not detected in the I band (DENIS survey) and it is not reported in other optical catalogues.
Combining the near and the mid-IR observations we have obtained the spectral infrared distribution for \source\/, shown in Fig.~\ref{figIR} (right panel) as open and filled circles. Using the reddening law derived by Rieke and Lebosky (1985), we have de-reddened the observed spectral points (triangles in Fig.~\ref{figIR}, right panel). This spectral energy distribution suggests the identification of \source\/ as a OB star.
In addition the de-reddened NIR colors of (J-H)$_0$=-0.11$\pm0.06$ and (H-K)$_0$=-0.12$\pm0.03$ are compatible with 
both OB Supergiant and OB main sequence, when compared with the colors for Supergiant and main sequence stars reported by Ducati et al. (2001).
Although IR spectroscopy of this source is needed to confirm this result, the X-ray behavior and infrared analysis strengthens the idea that  this source is a SFXT candidate.

\section{Discussion}

The INTEGRAL satellite has modified our understanding of high mass binary
systems, showing the existence of a new population of
compact objects orbiting
around a massive supergiant star and  exhibiting unusual properties, being either extremely absorbed,
or characterized by very short and intense flares.\\
The SFXTs are characterized by short and bright flares,
each with a duration between 10$^2$ s and 10$^4$ s, reaching 10$^{36-37}$ erg s$^{-1}$ at their peak.
In these transients, both the duration and shape of the bright flares are variable and 
the mechanism producing the transient emission is still not clear.
X-ray spectra (0.1--100 keV) during flares are well described either by a flat power law below
10 keV (Sidoli et al. 2006), with a photon index of 0-1 and a high energy
cut--off around 10--30 keV, or by a bremsstrahlung model with a temperature of 15--40 keV. 
A low intensity flaring, corresponding to an intermediate state between the quiescence and the flare peaks, is usually present, characterised by an X--ray luminosity
of 10$^{33-34}$ erg s$^{-1}$, as discovered thanks to XRT/\swift\/ monitoring campaign of SFXTs (Romano et al. 2010, 2011). In this state the X--ray spectrum is softer, with an absorbed power law
with photon index of 1--2 (Sidoli et al. 2008, Romano et al. 2010).
At the moment there are 11 firm established SFXTs and a similar number of candidates (Sidoli 2011).

\begin{table*}
\caption{Characteristics of \source\/ during the strongest outburst}
\tiny
\centering
\begin{tabular}{cccccccc}
\hline
\hline
\noalign{\smallskip} 
OBS. Tstart & Duration & Integrationa Time & $N_{\rm H}$ 		& $\Gamma$       	&  $F_{\rm unabs}$                   	& $Lum_{\rm 2-10keV}$ &  $\chi^2$/d.o.f. \\
\noalign{\smallskip}
         UTC& (ks)    	&(s)   & (10$^{22}$~cm$^{-2}$) &                	& (10$^{-12}$erg~cm$^{-2}$~s$^{-1}$)  	&          (erg s$^{-1}$)     &         \\
&&&&&&&\\
\hline
&&&&&&&\\
\noalign{\smallskip} 
1998-03-19  14:52:11.0  & 4.6& 1614.9 &18.9$^{+4.7}_{-3.8}$ & 2.0$\pm$0.5    	&  104$^{+50}_{-27}$ 				& 10$^{35}$  	&42/67		\\
&&&&&&&\\
\hline
\hline
\end{tabular}\\
\scriptsize
The continuum spectral model is an absorbed power law. The $N_{\rm H}$ is the absorption column density, 
$\Gamma$ is the power-law photon index, $F_{\rm unabs}$ is the unabsorbed flux in the 1.5-10~keV band. 
Outburst duration is the extrapolated activity period assuming as beginning of the outburst the first bin during which the source was detected while the integration time is the good time interval for the spectral extraction.\\
\normalsize
\label{tab:fit3}
\end{table*}

A time-resolved \sax\/ spectral analysis of \source\/ covering a significant range of orbital phases shows 
the source X--ray emission could be  well
described in the different selected time intervals 
by an absorbed power law model. 
While the photon index of the power law remained 
constant, the absorption column density was highly variable. We measured a significant rise in the 
$N_{\rm H}$ from $\sim$3 to 20$\times$10$^{22}$~cm$^{-2}$ across the transition 
from the low emission level (phase L) to the peak of the flare seen by SAX(phase G). We note that this does not correspond to the assumed periastron of the system based on the peak of the folded IBIS light curve but is likely associated with the accretion of a well-defined clump of wind.
Similar behavior is usually observed during the occurrence of X--ray eclipses 
in some of the sgHMXBs and SFXT (Bozzo et al. 2008, 2011). 
The neutron star could be obscured by its supergiant companion and for this reason   
the X--ray emission, produced by the accretion close to the surface of the NS,
is progressively absorbed in the photosphere of the supergiant star.

With this source being confirmed as an HMXB, its location is possibly associated to the spiral arms of our Galaxy, as usually observed with INTEGRAL 
HMXB population (Dean et al. (2005), Lutovinov et al. (2005), Bodaghee et al. (2007, 2011)). Assuming a distance of 3 kpc and 10 kpc for the spiral structure of the Galaxy,
the X-ray luminosity (2--10 keV) during the flare spans from 10$^{35}$ erg s$^{-1}$ to 10$^{36}$ erg s$^{-1}$,  respectively.
During the low emission level these values spans from 4$\times$10$^{33}$ erg s$^{-1}$ to 4$\times$10$^{34}$ erg s$^{-1}$. 
The X-ray characteristics observed with \sax\/ show the frequent micro activity typical of the intermediate state, with a luminosity value 
lying between the quiescence and the flare peaks, and a soft X--ray spectrum ($\Gamma\sim 2$).

Soon after the discovery of SFXT, to explain the bright and short outbursts, in't Zand (2005) proposed the sudden accretion of material from the clump wind of the supergiant; 
Negueruela et al. (2006a) suggested that SFXTs orbits should be highly eccentric to explain the low luminosity in quiescence. 
Later, the idea of a clump and spherically symmetric wind has been associated with an eccentric and/or wide orbit (Walter \& Zurita Heras 2007; Negueruela et al. 2007). 
Based on the shape of the light curve observed during the 2007 outburst from the SFXT IGR~J11215--5952 displaying periodic outbursts, Sidoli et al. (2007) proposed the presence of a second denser wind component in the form of an equatorial wind disk from the supergiant donor, inclined with respect to the orbital plane. 
The flaring activity was explained with a wind that is inhomogeneous and anisotropic.

All models predict that fast X-ray flares can be produced by 
sporadic capture and accretion onto the neutron stars hosted in the 
SFXTs. 

The wind accretion theory allow us to estimate the physical and geometrical parameters of the clumpy responsible of the transient emission.
The duration of the main observed flare during the \sax\/ observation of our source (phase G) is $\Delta t_{\rm flare}\sim$4.6~ks,  a typical time scale on which the neutron star crosses a dense clumpy. 
Neglecting the orbital velocity 
of the neutron star as reporting in L{\'e}pine et al. 2008, from the duration of the flare we obtain the radius of the clump accreted by the neutron star: 

\begin{equation}
R_{\rm cl} \simeq 1/2 v_{\rm w} \Delta t_{\rm flare} 
\end{equation} 

where $v_{\rm w}$ is the relative velocity between 
the clump and the neutron star.

During the time of the neutron stars periastron passage the absorption column density increase and it could be computed from the 
equation (4) given in Bozzo et al. 2011: 

\begin{equation}
N_{\rm H}\simeq 1.3\times10^{22} v_{\rm w8}^4 d_{\rm 3kpc}^2 {\rm cm^{-2}}
\label{eq:nh} 
\end{equation}

where $d_{\rm 3kpc}$ is the source distance in units of 3~kpc and $v_{\rm w8}$ is the wind velocity in $10^{8}~cm~s^{-1}$.\\
Using the $N_{\rm H}$ measured by the spectral analysis at the peak of the flare ($\sim$1.9$\times$10$^{23}$~cm$^{-2}$) and assuming a distance of 3 kpc,  Eq.~\ref{eq:nh} allow us to estimate a wind velocity of  $\sim$1900~km/s. 
This value agrees with the $\beta$-velocity law (Castor et al. 1975) which gives a typical velocity of 1000-2000 km/s for an OB supergiant and for a companion star losing mass in the form of a steady, homogeneous and spherically symmetric wind.\\ 
Assuming this wind velocity, we found the radius of the clump accreted by the neutron star $R_{\rm cl} \simeq 1/2 v_{\rm w} \Delta t_{\rm flare}\simeq4.4\times10^6$km.
In the framework of the Bondi-Hoyle-Lyttleton accretion theory, since only matter within a distance smaller than the accretion radius is accreted, the mass of the clumpy is

\begin{equation}
M_{\rm cl}=(R_{\rm cl}/R_{\rm acc})^2 M_{\rm acc}, 
\end{equation}

where $R_{\rm acc}$=2$G$$M_{\rm NS}$/$v_{\rm w}^2\simeq1.0\times10^5$km, 
$M_{\rm NS}$=1.4$M_{\odot}$ and $M_{\rm acc}$ is the mass accreted during the flare. 
The X-ray flux extrapolated in 0.1-10 keV energy range during the flare $F_{unabs}$=$L_{\rm X}$/(4$\pi$$d^2$)=($G$$M_{\rm NS}$$\dot{M}_{\rm acc}$/$R_{\rm NS}$)/(4$\pi$$d^2$) and the continuity equation $\dot{M}_{\rm acc}=\rho~v_{\rm w}\pi~R_{\rm acc}^2$ 
(where $d$ is the source distance and $R_{\rm NS}$=10~km 
is the neutron star radius) allowed to estimate the mass accreted onto the neutron star.
We derived $\dot{M}_{\rm acc}$=10$^{-8}M_{\odot}~{\rm year}^{-1~}$ (for a distance of 3 kpc) and M$_{\rm acc}\simeq2.5\times10^{19}~g$, thus
\begin{equation}
M_{\rm cl}=(R_{\rm cl}/R_{\rm acc})^2 M_{\rm acc}\simeq6\times10^{22} g
\end{equation}
Ducci et al. (2009) computed the expected clumpy characteristics in the framework of the Bondi-Hoyle accretion theory taking into account
the presence of clumps. The expected values of $M_{\rm cl}$ and $R_{\rm cl}$ according to this clumpy wind model are in agreement with the measured values for \source\/. We conclude that \source\/ can be added to the list of Supergiant Fast X-ray Transients as it shows all of the characteristics typical of that class of object.

\acknowledgments
The authors acknowledge the ASI financial support
via ASI/INAF grants I/033/10/0.

\scriptsize
\section{References}
 Bird et al. 2010, \apjs, 186,1\\
Baumgartner et al. 2010, HEAD, 11, 1305.\\
Bodaghee et al. 2007, A\&A, 467, 585\\
 Bodaghee et al. 2011, ApJ in press\\
Boella et al. 1997, A\&AS, 122, 299\\
 Bozzo, E., Falanga, M., \& Stella, L. 2008, \apj, 683, 1031\\
 Bozzo et al. 2011, A\&A, 531, 130\\
 Burrows, D. N. et al., 2005, Space Sci. Rev., 120, 165\\
 Chaty s. 2008, ChJGAS, 8, 197\\
 Clark et~al., 2009, \mnras, 399, 113\\
 Clark et~al., 2010, \mnras, 406, 75\\
 Corbet et al. 2010, ATel \# 2588\\
 Courvoisier, T. J.-L., et al., 2003, \aap,  411, L53\\
 Cusumano et al. 2010, \aap, 510, 48\\
 Dean et al. 2005, A\&A, 443, 485\\
 Drave et~al., 2010, \mnras, 409, 1220\\
 Ducati J. R., Bevilacqua C. M., Rembold S. B. et al. 2001, \apj, 558, 309\\
 Ducci, L., Sidoli, L., Mereghetti, S., Paizis, A., \& Romano, P. 2009, \mnras, 398, 2152\\
 Federici M., Martino B. L. and Natalucci L., 2009 PoS2009-092\\
  Filliatre and Chaty, 2004, ApJ, 616, 469\\
  Fiocchi et al. 2010, ApJ, 725, 68\\
 Fiore, F., Guainazzi, M., \& Grandi, P. 1999, Cookbook for BeppoSAX NFI Spectral Analysis (www.asdc.asi.it/bepposax/software/cookbook)\\
 Frontera et al, 1997, A\&AS, 122, 357\\
 Grupe et al. 2009, ATel \# 2075\\
Hill et~al., 2004, Proc. SPIE, 5165, 217\\
 in't Zand, J. J. M. 2005, A\&A, 441, L1\\
 Ibarra et al., 2007, A\&A ,465, 501\\
 Lepine, S. \& Moffat, A.~F.~J. 2008, \aj, 136, 548\\
 Lomb, N. R. 1976, Ap\&SS, 39, 447\\
 Lutovinov, A., Revnivtsev, M., Molkov, S., et al. 2005, A\&A, 430, 997\\
 Manzo et al. 1997, A\&AS, 122, 341M\\
 Negueruela, I., Smith, D. M., Harrison, T. E., \& Torrejón, J. M. 2006a, ApJ, 638, 982\\
 Negueruela et al. 2006b, ESASP, 604, 165\\
 Negueruela, I., Smith, D. M., Torrejon, J. M., \& Reig, P. 2007, astro-ph/0704.3224\\
Negueruela, I., Torrejon, J. M., Reig, P., Ribo, M., \& Smith, D. M. 2008, arXiv:0801.3863\\
 Parmar et al. 1997, A\&AS, 122, 309\\
 Rieke and Lebosky, 1985, AJ,288,618\\
 Romano et al. 2010, MNRAS, 401, 1564\\
 Romano et al. 2011, 2011 Fermi Symposium proceedings - eConf C110509, astro-ph/1111.0698\\
 Scargle, J. D. 1982, ApJ, 263, 835\\
 Sidoli, L., Paizis, A., \& Mereghetti, S., 2006, A\&A, 450, L9\\
 Sidoli, L., Romano, P., Mereghetti, et al. 2007, A\&A, 476, 1307\\
 Sidoli, L.; Romano, P.; Mangano, V., et a., 2008, ApJ, 687, 1230\\
 Sidoli et al. 2009,  MNRAS, 397, 1528\\
 Sidoli et al. 2010,  ApJ, 690, 120\\
 Sidoli L., 2011,   astro-ph/1111.5747\\
 Smith et al. 2006, \apj, 638, 974\\
 Sguera et al. 2005, \aap, 444, 221\\
 Sguera et al. 2006, \aap, 646, 452\\
 Sguera et al. 2007, \aap, 492, 163\\
Sidoli L., Paizis A. and Mereghetti S., 2006, \aap,  450, L9\\
 Ubertini, P., et al. 2003, \aap,  411, L131\\
 Winkler, C. et al. 2003, \aap, 411, 349\\
Walter, R., \& Zurita Heras, J. 2007, A\&A, 476, 335\\


\end{document}